%% file: ms.tex
\documentclass[12pt,preprint]{aastex}

\shorttitle{GRB~051111 KAIT and LOT}
\shortauthors{Butler et al.}

\def\gtrsim{\mathrel{\hbox{\rlap{\hbox{\lower4pt\hbox{$\sim$}}}\hbox{$>$}}}}
\def\lessim{\mathrel{\hbox{\rlap{\hbox{\lower4pt\hbox{$\sim$}}}\hbox{$<$}}}}

\begin{document}

\title{When Do Internal Shocks End and External Shocks Begin?
Early-Time Broadband Modelling of GRB~051111}

\author{N.~R. Butler\altaffilmark{1,2},
W. Li\altaffilmark{2},
D. Perley\altaffilmark{2},
K.~Y. Huang\altaffilmark{5},
Y. Urata\altaffilmark{6,8},
J.~X. Prochaska\altaffilmark{3},
J.~S. Bloom\altaffilmark{2},
A.~V. Filippenko\altaffilmark{2},
R.~J. Foley\altaffilmark{2},
D. Kocevski\altaffilmark{2},
H.-W. Chen\altaffilmark{4},
Y. Qiu\altaffilmark{7}, 
P.~H. Kuo\altaffilmark{5}, 
F.~Y. Huang\altaffilmark{5},
W.~H. Ip\altaffilmark{5},
T. Tamagawa\altaffilmark{6},
K. Onda\altaffilmark{8},
M. Tashiro\altaffilmark{8},
K. Makishima\altaffilmark{6,9},
S. Nishihara\altaffilmark{10},
and Y. Sarugaku\altaffilmark{10}}

\altaffiltext{1}{Townes Fellow, Space Sciences Laboratory, University of California,
Berkeley, CA, 94720-7450}
\altaffiltext{2}{Astronomy Department, University of California,
445 Campbell Hall, Berkeley, CA 94720-3411, USA}
\altaffiltext{3}{Department of Astronomy and Astrophysics, 
University of California, Santa Cruz, CA 95064, USA}
\altaffiltext{4}{Department of Astronomy and Astrophysics, University
of Chicago, 5640 South Ellis Avenue, Chicago, IL 60637, USA}
\altaffiltext{5}{Institute of Astronomy, National Central University, Chung-Li
   32054, Taiwan, Republic of China}
\altaffiltext{6}{RIKEN (Institute of Physical and Chemical Research), 2-1 Hirosawa,
   Wako, Saitama 351-0198, Japan}
\altaffiltext{7}{National Astronomical Observatories, Chinese Academy of Sciences,
   Beijing 100012, China}
\altaffiltext{8}{Department of Physics, Saitama University, Shimo-Okubo, Sakura,
   Saimtama, 338-8570, Japan}
\altaffiltext{9}{Department of Physics, University of Tokyo, 7-3-1 Hongo, Bunkyo-ku,
   Tokyo 113-0033, Japan}
\altaffiltext{10}{ISAS/JAXA, 3-1-1 Yoshinodai, Sagamihara, Kanagawa, 229-8510, Japan}

\begin{abstract}
Even with the renaissance in gamma-ray burst (GRB) research fostered by
the Swift satellite, few bursts have both contemporaneous observations
at long wavelengths and exquisite observations at later times across
the electromagnetic spectrum.  We present here contemporaneous imaging
with the KAIT robotic optical telescope, dense optical sampling with
Lulin, and supplemented with infrared data from PAIRITEL and radio to
gamma-ray data from the literature.  For the first time, we can test
the constancy of microphysical parameters in the internal-external
shock paradigm and carefully trace the flow of energy from the GRB to
the surrounding medium.  KAIT data taken $\lessim 1$ minute after the
start of GRB~051111 and coinciding with the fading gamma-ray tail of
the prompt emission indicate a smooth re-injection of energy into the
shock.  No color change is apparent in observations beginning $\sim 1.5$
minutes after the GRB and lasting for the first hour after the burst.
There are achromatic flux modulations about the best-fit model at late 
($t\approx 10^4$ s) times, possibly due to variations in the external density.
We find that the host-galaxy extinction is well fit by a curve similar to that of
the Small Magellanic Cloud.  Low visual extinction, $A_V\approx 0.2$
mag, combined with high column densities determined from the X-ray
and optical spectroscopy ($N_H> 10^{21}$ cm$^{-2}$), indicate a low
dust-to-metals ratio and a possible over-abundance of the light metals.
An apparent small ratio of total to selective extinction ($R_V\approx 2$)
argues against dust destruction by the GRB.  Time constancy of
both the IR/optical/UV spectral energy distribution and the soft X-ray
absorption suggests that the absorbing material is not local to the GRB.
\end{abstract}


\maketitle

\section{Introduction}

The {\it Swift}~gamma-ray burst (GRB) satellite \citep{gehrels04} continues to unleash a torrent
of finely time- and energy-sampled photons arising from the bursts and their impact on
the surrounding medium (i.e., the GRB afterglows).
The X-ray Telescope (XRT) provides detailed light curves for 1-2 bursts and burst
afterglows per week, allowing us to routinely view the practically uncharted first 100s to $\sim 1$ day 
in the life of a GRB.  In the X-ray data, we see complex phenomena:
unexpected rapid decays, also sometimes temporally flat and apparently re-energized
afterglows, and massive X-ray flares \citep[see, e.g.,][]{nousek06}.  Unfortunately, it has proven
very challenging to complement these data with an early and rapid-cadence data set 
at longer wavelengths.  Because studies of GRB afterglows prior
to {\it Swift}~rely primarily on observations in the optical and longer wavelengths, and mostly
at later times, contemporaneous
observations are critical for connecting new-found insights to the large body of previous
work.  Correlated, broadband 
observations of events which exhibit extremely energetic broadband emission from
the radio to the gamma-rays (e.g., Figure \ref{fig:broadband_spec}) allow us to study and
potentially understand the new phenomenology and to pose answers to numerous open questions.

Optical observation of GRBs in the prompt phase and shortly there after
are extremely rare.  Of $\sim 50$ optical afterglows detected prior to {\it Swift},
only one \citep[GRB~990123,][]{akerlof99} was detected during the prompt phase and
only a handful were detected in the first 10 min of the afterglow:
GRB~021004 \citep{fox03a}, GRB~021211 \citep{fox03b,li2003b,vestrand04}, and GRB~030418 \citep{rykoff04}.
In the {\it Swift}~era, the rate of early detections has jumped dramatically, thanks
to rapidly communicated and tight GRB localizations from {\it Swift}~and also due to
the maturing system of ground-based robotic telescopes (e.g., $\sim$15 early
detections by ROTSE\footnote{http://www.rotse.net} alone).
The UV Optical Telescope (UVOT) on {\it Swift}~has the potential to match this performance,
particularly with a recent prioritization of early unfiltered observations
of the very red afterglows.
Perhaps most impressive, the RAPTOR experiment has detected two GRBs during the
prompt phase: GRB~041219 \citep{vestrand05} and GRB~050820A, \citep{vestrand06}.
The prompt and contemporaneous long wavelength emission of GRB~041219A was 
actually discovered at IR wavelengths \citep{blake05} with
the Peters Automated Infrared Imaging Telescope \citep{bloom06}.
Even though the Galactic extinction
toward the GRB was large, GRB~041219A is the only prior burst  
with prompt, long-wavelength observations at
multiple frequencies.

Here we present
observations of GRB~051111 conducted with the robotic 0.76-m Katzman Automatic Imaging Telescope
\citep[KAIT;][]{fil01,li2003a,li2003b}, beginning with an unfiltered exposure 43.7 s after the 
GRB trigger from the Burst
Alert Telescope (BAT) on {\it Swift}~and just catching the tail of the prompt emission.  
We obtained color information in the form of $V$ and $I$-band observations beginning just 73.7 s after
the trigger.  Supplemented with $B$, $V$, $R$, and $I$-band observations with the Lulin One-meter Telescope and $J$, $H$, and
$K_s$-band observations
with PAIRITEL taken hours after the burst and with other observations reported in the
literature, we present an unrivaled and impressive broadband view of this afterglow and its
early evolution.

The study of the early emission during and following a GRB  
is a critical step toward understanding the
            origin of the emission and the nature of the surrounding  
medium.
It is widely accepted that GRBs are produced by a self-interacting relativistic outflow \citep[``internal
shocks,'' e.g.,][]{fmn96} which heats and shocks the surrounding medium (``external shocks'') to form a long-lasting
afterglow at longer wavelengths \citep{mnr97,snp99}.
Transient emission from a ``reverse shock,'' which propagates backward toward the central engine in the
shock frame, can also be produced.  Despite an early indication to the contrary from GRB~990123 \citep[e.g.,][]{akerlof99}, the
reverse shock does not appear to be a common feature in the early optical data \citep[see, also,][]{mkp06}.
Primarily due to the larger physical size, the external shocks are expected to generate smoother light
curves, and the $\sim 10$\% of GRBs with smooth time histories may require only external shocks \citep[e.g.,][]{mkp04}.

However, the growing sample of early afterglow observations militates against such a simple interpretation.  As mentioned
above, the X-ray data \citep[and also some optical data, e.g.,][]{fox03a,woz05} evidence shock refreshment 
\citep{reesNmes98,snm00,rretal01} or continued central 
engine activity \citep{reesNmes00,macfad01,rr04,leeNrr02},
which can persist for several hundreds of seconds or longer \citep[e.g.,][]{nousek06}.  
While its physical size is still small compared to the timescale for synchrotron cooling, the external shock can cool rapidly.  
Energy can be efficiently 
radiated from the shock at this stage \citep[e.g.,][]{spn98}, prior to the onset of an adiabatic evolution.  Depending on the breadth of the
shock shell, which depends on the duration of the period of internal shocks generation, the internal shock emission can overlap with
the external shock emission.  
Below, we use the exquisitely sampled data for GRB~051111 to disentangle the early-time emission components
and also to study the dust and gas properties of the host galaxy.

\section{Observations and Data Reduction}
\label{sec:redux}

At 05:59:41.4 UT, {\it Swift}~BAT triggered and localized GRB~051111 \citep{sakamoto05}.
The 3\arcmin~radius error region from BAT enabled the rapid detection of a new optical source only 26.9s
after the trigger \citep{rujo05}, while the burst was still in progress.  
A barrage of detections in the optical, ultraviolet (UV),
infrared (IR), radio, and X-ray bands followed.   Figure 1 shows a snapshot of the full data set,
interpolated via the modelling below to $t=100$ s.

\subsection{Gamma- and X-ray}

We downloaded the {\it Swift}~BAT gamma-ray data from the {\it Swift}~Archive\footnote{ftp://legacy.gsfc.nasa.gov/swift/data}.  The
energy scale and mask weighting were established by running the {\tt
bateconvert} and {\tt batmaskwtevt} tasks from HEAsoft~6.0.4.  Spectra
and light curves were extracted with the {\tt batbinevt} task, and response
matrices were produced by running {\tt batdrmgen}.  We apply the systematic
error corrections to the low-energy BAT spectral data as suggested by the BAT
Digest website\footnote{http://swift.gsfc.nasa.gov/docs/swift/analysis/bat\_digest.html}, and fit the data using 
ISIS\footnote{http://space.mit.edu/CXC/ISIS/}.

The {\it Swift}~XRT X-ray data from 5 follow-up observations of the GRB~051111 field
were downloaded from the {\it Swift}~Archive and reduced by 
running version 0.10.3 of the {\tt xrtpipeline} reduction script from the 
HEAsoft~6.0.6\footnote{http://heasarc.gsfc.nasa.gov/docs/software/lheasoft/}
software release.  From there, we bin 
the data in time, exclude pileup
chip regions for each time interval, account for lost flux due to bad pixels,
and produce spectra using custom IDL 
scripts \citep[e.g.,][]{butler06}.  The data cover the time range from 5.56 
ksec until 405.29 ksec after
the burst, with a total exposure (livetime) of 51.98 ksec.  
Spectral response files are generated using the {\tt xrtmkarf} task and
the latest calibration database (CALDB) files (version 8, 2006-04-27).
The spectra are modelled using ISIS.
For each spectral bin, we require a signal-to-noise ratio (S/N) of 3.5. 
We define S/N as the background-subtracted number of
counts divided by the square root of the sum of the signal counts and the
variance in the background.  We define the background region as that 
where the number of counts in an aperture the size of the source extraction
region is within $2\sigma$ of the median background over the chip
in that aperture for one contiguous follow-up observation.  For the 
Photon Count (PC) mode data, the source aperture is a circle of 
radius 16 pixels.

\subsection{Optical}

The robotic 0.76-m Katzman Automatic Imaging Telescope
\citep[KAIT;]{fil01,li2003a,li2003b}.
at Lick Observatory observed GRB~051111 in a series of images
automatically obtained starting at 05:60:25.2 UT
\citep[43.7 s after the BAT trigger;][]{li2005}. 
The sequence includes a combination of
images taken with the $V$ and $I$ filters, as well as some that
are unfiltered (Table 1). The optical transient first identified by \citet{rujo05}
at $\alpha =23^h 12^m 33.2^s$,  $\delta =+18^{\circ} 22' 29.1''$ (J2000.0)
was clearly detected in each exposure.
We began $B$, $V$, $R$, and $I$ band imaging \citep{huang05} of the
GRB~051111 afterglow using
the Lulin One-meter Telescope \citep[LOT;][]{huang05b,urata05} at 10:07 UT.
We detected the optical afterglow
clearly in each band until observations ceased at 14:42:06 UT (Table 2).

We find that the combination of the KAIT optics and the quantum efficiency
of the Apogee CCD camera makes the KAIT unfiltered observations mostly
mimic the $R$ band.  We determine a small color correction \citep{li2003a,li2003b,riess99}
between the unfiltered and $R$ band photometry
of $0.11\pm 0.02$ mag using the $(V-R)$ and $(R-I)$ color information from
the Lulin observations.

We use point spread function (PSF) fitting through the IDL DAOPHOT\footnote{http://idlastro.gsfc.nasa.gov}
package to reduce our data.  Only
unsaturated and spherically symmetric sources within a given CCD exposure are
used to model the stellar PSF for that exposure.
The absolute photometric calibration for KAIT is determined using
more than 10 \citet{lan92} fields observed on Nov. 24 UT at a range of airmasses.
The Lulin photometry is calibrated using \citet{lan92} fields
PG0231+051, SA92, and SA97, and also cross-checked using the KAIT
standard star observations.

\subsection{Near-IR and UV}

After initially poor transmission conditions (i.e., clouds) at Mt. Hopkins,
we obtained a total imaging exposure of 188 s on the field of
GRB~051111 before hitting a telescope limit. In simultaneous observations
with PAIRITEL \citep{bloom06} we measure magnitudes of the \citet{rujo05} transient
of $J = 16.55 \pm 0.03$, $H = 15.85 \pm 0.04$, and $K_s = 15.29 \pm  0.06$ \citep{bloom05},
relative to 2MASS (Two Micron All Sky Survey).  The data were taken
from 07:23:29 to 07:28:26 UT, centered 1.44 hrs after the BAT trigger.

We refine the {\it Swift}~UVOT photometry \citep{poole05} in the $V$, $B$, 
and $U$
bands by employing a tight extraction region in order to maximize the S/N
\citep[see][]{li2006}.  We find $V=19.38 \pm 0.15$ 
($t=16298-17198$ s post BAT trigger), $B=20.04 \pm 0.10$ ($t=
11237-12136$ s post BAT trigger), and $U=19.81 \pm 0.12$ ($t=
10329-11229$ s post BAT trigger).  These numbers are consistent with those
of \citet{poole05}, but with considerably smaller error bars.
We extrapolate the Galactic extinction curve \citep{sfd98}
into the UV in order to perform the fitting for the UVW1, UVM2,
and UVW2 bands below.

\section{Fits to the Data}

\subsection{Optical Light Curve and Lack of Color Change}
\label{sec:nocolor}

\begin{figure}
\centerline{\includegraphics[width=7.0in]{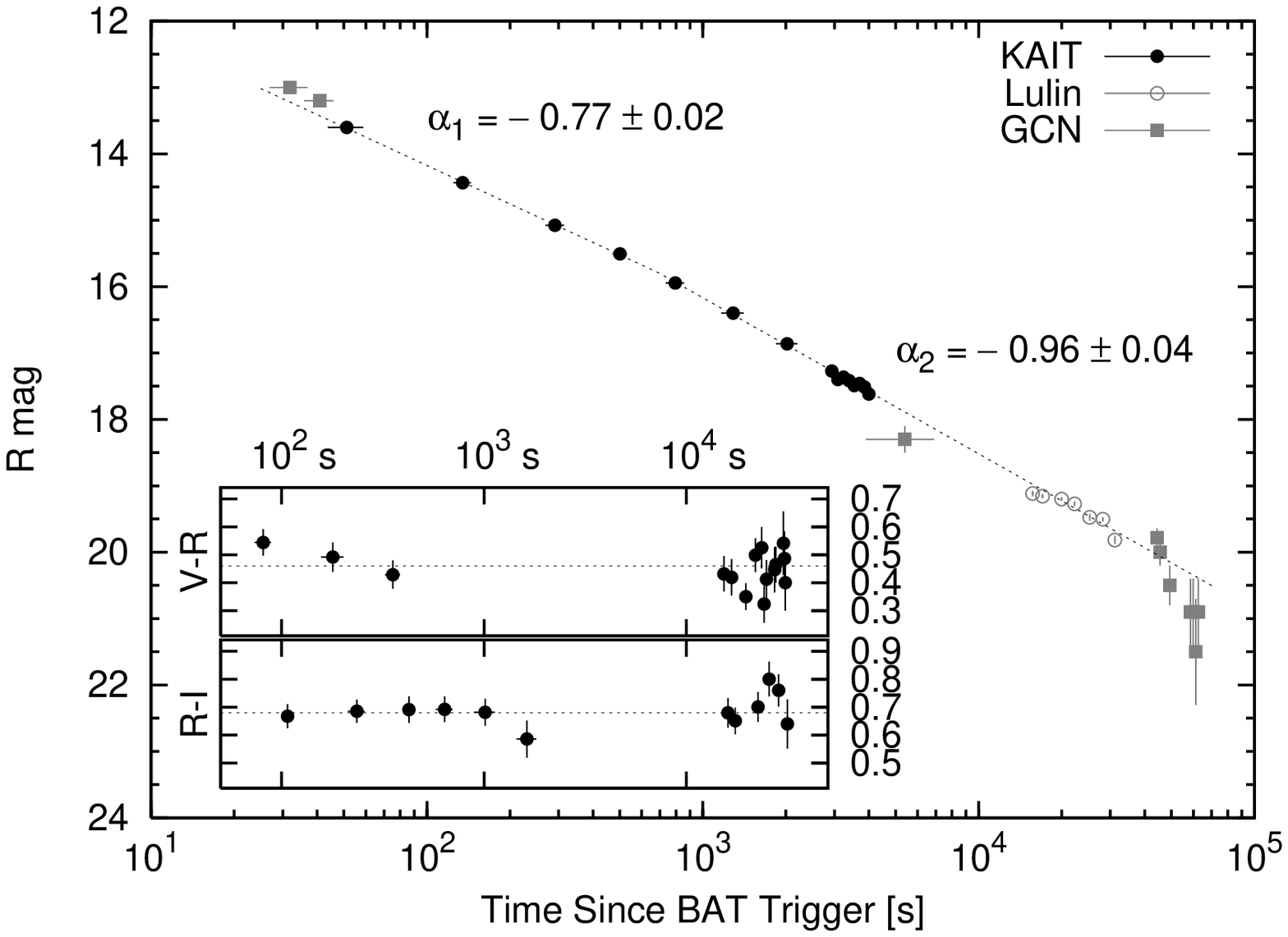}}
\caption{\small $R$ band light curve and $(V-R)$, $(R-I)$ colors
from KAIT and Lulin.  The best-fit temporal decay indices ($\alpha$) for the source flux are also
shown.  The unfiltered and $R$ band data are from the GCN 
\citep{garimella05, milne05, nanni05, rujo05, ryko05, sharapov05, smith05}.}
\label{fig:lc_opt}
\end{figure}

As shown in Figure \ref{fig:lc_opt}, the KAIT $R$ band data are well fit by a broken
powerlaw with $t_{\rm break} = (700\pm 260)$ s ($\chi^2/\nu=5.61/11$).  The best-fit flux
decay indices are shown in Figure \ref{fig:lc_opt}.  The
fit is statistically unacceptable without the break ($\chi^2/\nu=65.15/13$, $\alpha=0.83\pm0.01$).
Including the Lulin $R$ band data in the fit, the model parameters do not change, but
the fit quality degrades ($\chi^2/\nu=38.9/20$).  Similar fit qualities are found
for the data in the $B$, $V$, and $I$ bands by applying magnitude offsets to the best-fit $R$ band
model ($\chi^2/\nu= 27.0/11$, $54.1/13$, and $64.0/10$, respectively).  From the fits, we derive 
the following afterglow colors: $(B-R)= 0.97 \pm 0.02$, $(V-R)= 0.37 \pm 0.01$, and $(R-I)= 0.80 \pm 0.01$ mag,
corrected for Galactic extinction \citep{sfd98}.

\begin{figure}
\centerline{\includegraphics[width=7.0in]{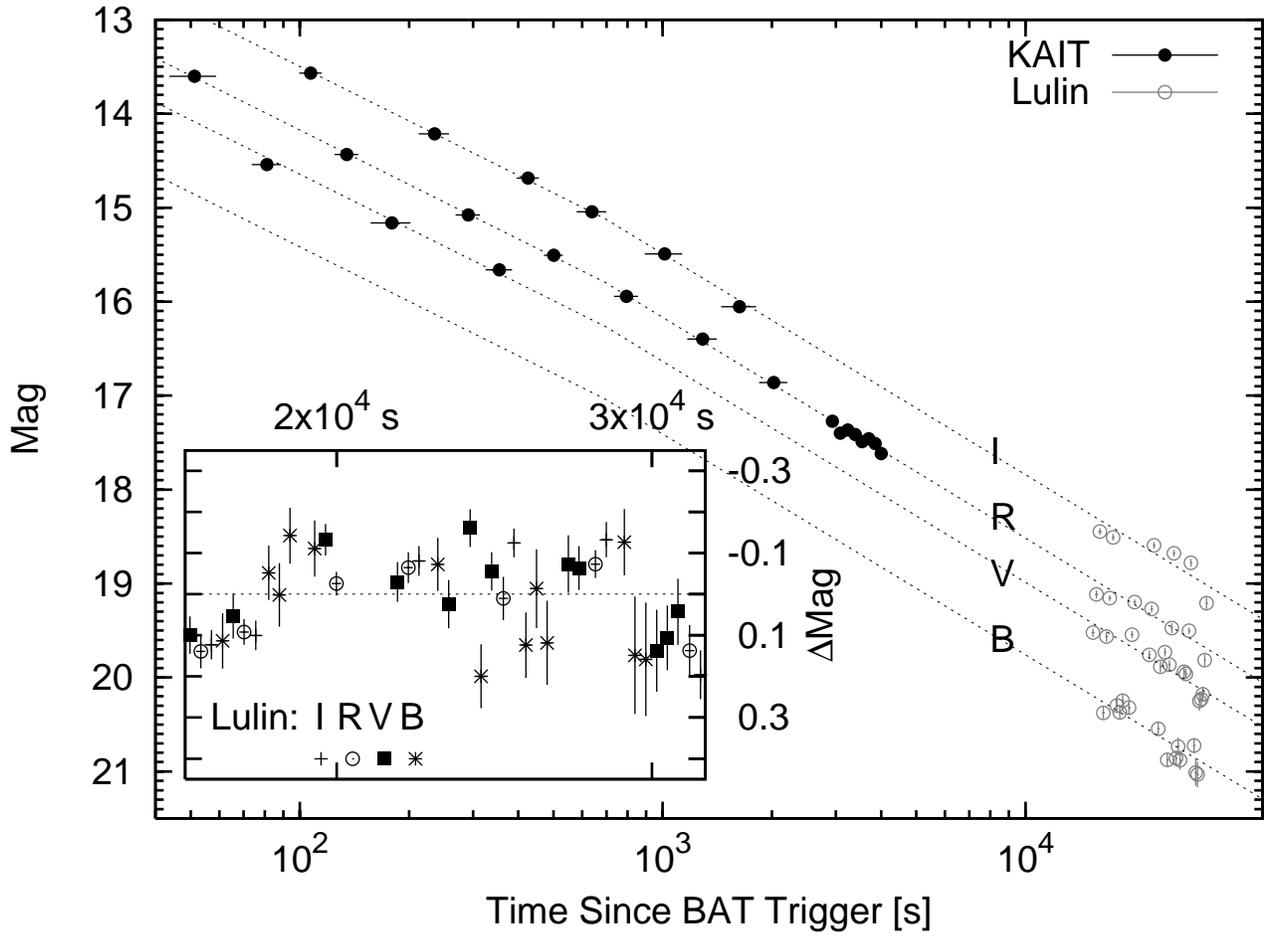}}
\caption{\small Multi-color light curve of GRB~051111 from KAIT and Lulin.  The fit
residuals for the Lulin data are plotted in the inset.}
\label{fig:lc_opt2}
\end{figure}

\begin{figure}
\centerline{\includegraphics[width=7.0in]{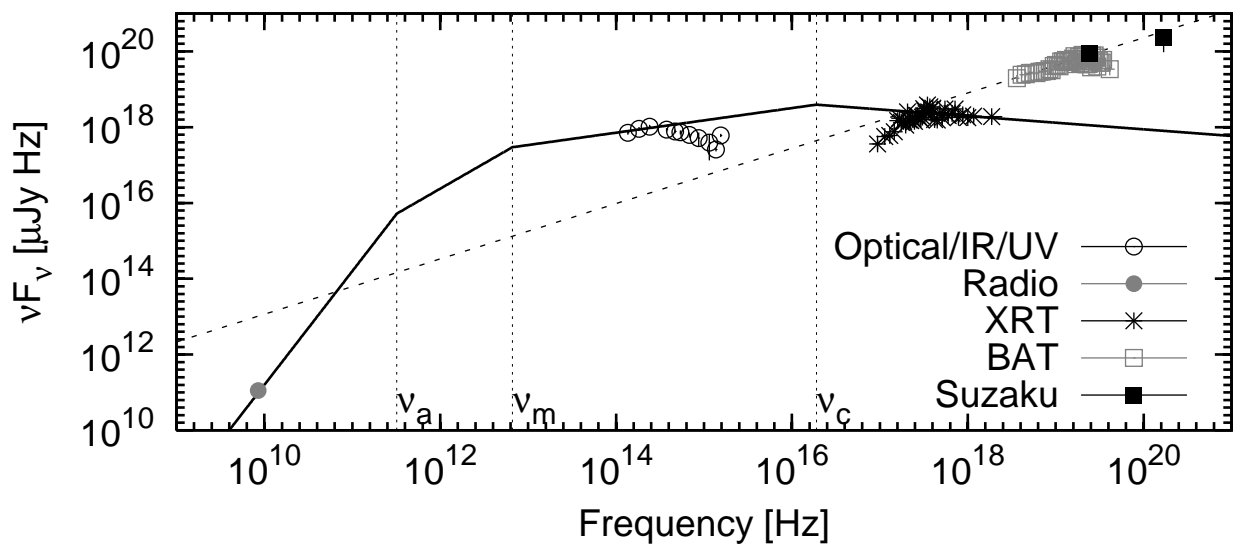}}
\caption{\small Broadband model overview, data shown at $t=100$ s.  The afterglow data in the radio, optical, IR, UV, and X-rays
are well fit by a external shock expanding into a uniform-density medium and emitting synchrotron radiation
(Sections \ref{sec:broadband1}, \ref{sec:broadband2}).
The optical and X-ray data drop away from the model curve at high and low energies, respectively,
due to an appreciable amount of absorption by dust and metals (respectively).  Also shown are the prompt
gamma-ray spectra from {\it Swift}~BAT and {\it Suzaku}~WAM.}
\label{fig:broadband_spec}
\end{figure}

The $(R-I)$ and $(V-R)$ colors, without the Galactic extinction correction, are plotted in the sub-panel of 
Figure \ref{fig:lc_opt}, and the multi-band
data are plotted in Figure \ref{fig:lc_opt2}, along with the fits.  The KAIT and Lulin data are consistent with
no color evolution, and the light-curve break is most likely achromatic.
From the $(V-I)$ colors before and after the break, we derive a change in the spectral index $\delta \beta < 0.38$ ($3\sigma$).
Figure \ref{fig:lc_opt2} displays another interesting characteristic of the data: there
are strong residuals in the Lulin data with $\delta t/t\approx 0.1$ relative to the
broken powerlaw fit, and these are correlated across spectral bands.

As we discuss
further below, the PAIRITEL data, as well as the available data from the GCN in
the optical and UV, are consistent with the broken powerlaw decay, modified
by dust absorption.

\subsection{Gamma- and X-ray}
\label{sec:xandg}

The BAT gamma-ray light curve exhibits a characteristic FRED-like (``fast rise exponential decline'') time decay, with duration $T_{\rm 90}\approx 30$ s.
From a wavelet analysis, we determine the start of the burst as 7.17 s prior to the BAT trigger (Figure \ref{fig:broadband_lc2}).
From this point onward, we use this time as the beginning of the burst.  This has the effect of increasing the initial optical decay
index from 0.77 to 0.81 (Figure \ref{fig:lc_opt}), but does not affect any of the other results above.
The 15--150 keV spectrum from the beginning of the burst to 40 s later is adequately fit by a powerlaw with photon index $\Gamma=1.27\pm0.05$
($\chi^2/\nu=53.65/51$).  The model fluence in the 15--100 keV band
is $(2.1 \pm 0.1) \times 10^{-6}$ erg cm$^{-2}$.  The fits are consistent with those reported
by \citet{sakamoto05} and \citet{krimm05}, where a modestly longer burst time extraction region is applied.

There is a significant amount of spectral evolution, comparing the time prior to burst peak to the time post peak.
For the burst rise from 0 s to 10 s, we find $\Gamma_1=1.11 \pm 0.07$ and
$F_{\rm 15 keV}=(0.48\pm 0.04)$ mJy ($\chi^2/\nu=49.35/51$).  For the burst decline from 10 s to 40 s, we find $\Gamma_2=1.38 \pm 0.08$ and
$F_{\rm 15 keV}=(0.31\pm 0.03)$ mJy ($\chi^2/\nu=41.15/45$).
The photon index for 20 s to 40 s ($\Gamma_3=1.4\pm 0.1$, $\chi^2/\nu=29.28/31$) is 
consistent with the 10 s to 40 s value, indicating that this value provides a 
reasonable characterization for the full decline phase.
The burst was also observed by the {\it Suzaku}~Wideband All-sky Monitor \citep[WAM;][]{yamaoka06} 
and found to have a similar time profile as from {\it Swift}~and also
a consistent time integrated spectral slope
($\Gamma=1.5\pm 0.3$) and a fluence in the 100--700 keV band of $(8.4 \pm 0.8) \times 10^{-6}$ erg cm$^{-2}$ \citep{yamaoka05}.  Combining the fluence 
determinations from {\it Swift}~and {\it Suzaku},
we find a 15--700 keV fluence of $(1.05\pm 0.08) \times 10^{-5}$ erg cm$^{-2}$.
Because the peak $\nu F_{\nu}$ energy for this burst has not been measured, the bolometric fluence could be substantially
larger than the 15--700 keV fluence.  Assuming that all the prompt photons have been accounted for, a lower limit on the
isotropic equivalent energy emitted in the host frame at $z=1.55$ \citep{hill05} is $E_{\rm iso}=6.2 \times 10^{52}$ erg.  
Here and throughout, we assume a cosmology with ($h$, $\Omega_m$, $\Omega_{\Lambda}$) $=$ (0.71, 0.3, 0.7).

The X-ray light curve in the 0.3--10.0 keV band is reasonably well fit ($\chi^2/\nu=38.7/32$) by a powerlaw
model
with $\alpha=-1.6\pm0.1$.  The spectrum in the same band contains 573 counts and is well-fit ($\chi^2/\nu=43.61/36$) by an absorbed powerlaw
with photon index $\Gamma = 2.3\pm0.2$, absorbing column $N_H = 1.8\pm0.4 \times 10^{21}$ cm$^{-2}$,
and unabsorbed flux $(5.0\pm 0.5) \times 10^{-13}$ erg cm$^{-2}$ s$^{-1}$, for the time period 5.56 to
405.29 ksec after the GRB.  We use the corresponding average spectral flux at 1 keV, $(6.6\pm 0.9) \times 10^{-2}$ $\mu$Jy, to translate the XRT 
count rate to $\mu$Jy below.
The absorption is greater than the inferred Galactic value in the source
direction \citep[$N_{H,{\rm Galactic}} = 5 \times 10^{20}$ cm$^{-2}$,][]{dickey1990} at $4.1\sigma$
significance ($\Delta \chi^2=16.48$, for 1 additional degree of freedom).
At redshift $z=1.55$ \citep{hill05}, the excess column is $(7\pm 4)\times 10^{21}$ cm$^{-2}$, allowing for a 20\% uncertainty in
the Galactic column.  We find no evidence for $\gtrsim 1\sigma$
significant emission lines in the spectrum.  These fits are consistent with the preliminary fits reported by \citet{parola05}.

\section{Discussion}

\subsection{Late-Time Broadband Afterglow Modelling}
\label{sec:broadband1}

\begin{figure}
\centerline{\includegraphics[width=7.0in]{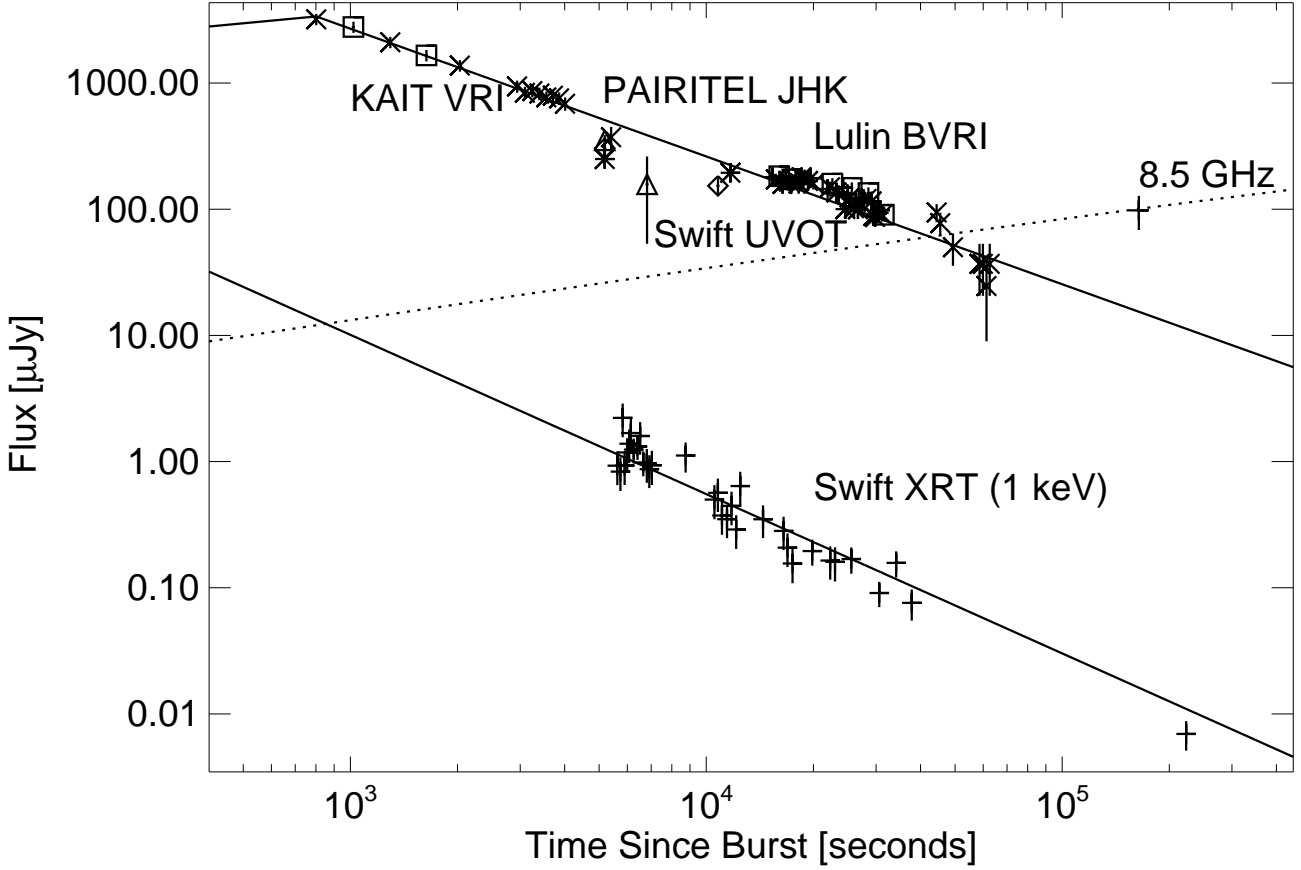}}
\caption{\small The data for $t\gtrsim 700$ s are well fit by a synchrotron external shock model.  
The optical, UV, and IR data are scaled to $R$ band using the best-fit synchrotron
shock plus absorption model (Table 3, $A_{R{\rm ~obs}}=0.72$ mag, Figure \ref{fig:sed}).  All data include a 10\% systematic error
added in quadrature to the statistical error.  The dip near $10^4$ s is seen in the $R$ band data as well
as the UVOT data and the data from PAIRITEL (Section \ref{sec:sed}).  We include the GCN data
\citep{frail05, garimella05, milne05, nanni05, poole05, rujo05, ryko05, sharapov05, smith05},
in addition to the data from KAIT, PAIRITEL, Lulin, and {\it Swift}. All data are corrected for Galactic
extinction \citep{sfd98}.}
\label{fig:broadband_lc}
\end{figure}

After $t\approx 10^3$ s, the broadband afterglow data are well described by
the external shock synchrotron model \citep{pnr93,katz94,waxman97,wrm98}.
We do not attempt to model the slowly decaying early optical light curve
or the achromatic (or nearly so) optical break discussed above, nor do we
fit for the BAT data or the apparent fluctuations in the optical at
$t\approx 6000$ s (Figure \ref{fig:lc_opt2}) at this stage.   These features are discussed below.
Due to the fluctuations in the optical about the best fit model at $t\sim 10^4$ s,
we are also forced to add a 10\% systematic
uncertainty component in quadrature to the statistical uncertainty of each
data point, in order to obtain an acceptable fit ($\chi^2/\nu=118.1/103$,
$t\ge 700$ s).  
Figure \ref{fig:broadband_lc} shows the best-fit model, plotted over the radio, optical,
UV, IR, and X-ray data at $t\ge 700$ s.  In the next section, we describe the modelling 
of the host-frame extinction by dust, which has been taken into account (Figure \ref{fig:sed}) 
in the figure so that
the data in the optical, UV, and IR appear on one common curve.

Because the optical time decay is shallow and the X-ray decay is steep, we
expect and observe the data to be well fit
by a uniform density, interstellar medium (ISM) model.  A model with $n\propto R^{-2}$, describing
the expected density contours for a progenitor star with a significant
wind, would exhibit a synchrotron cooling break frequency $\nu_c$ which increases
in frequency with time \citep{chevNli00}.  This leads to shallower decays at high
versus low energy, opposite to the behavior of the ISM model.
For the constant-density model, the optical and X-ray decay indices ($\alpha_{\rm
opt}=-0.96\pm 0.04$, $\alpha_{\rm X}=-1.6\pm 0.1$) and the X-ray energy
index ($\beta_{\rm X}=-1.3\pm 0.2$)
constrain the powerlaw index $p$ describing the energy distribution of
synchrotron-emitting electrons: $p=2.35\pm 0.05$ \citep[e.g.,][]{spn98}.  
In the best-fit model, the synchrotron cooling break lies between the optical and X-ray bands
during the observation.  The predicted value for the optical
energy index is $\beta_{\rm opt}=-0.68\pm 0.03$.  

The model contains 4 additional free
parameters \citep[e.g.,][]{spn98}: $\eta_{\gamma}$, $\epsilon_e$, $\epsilon_B$, and $n$, the
efficiency factor relating the shock energy to the energy released in
gamma-rays, the
fractions of equipartition energy going into the electrons and magnetic
field $B$, and the density $n$.  These are constrained by the three
observed
radio, optical, and X-ray fluxes and by the equation describing the time
evolution of the synchrotron spectral peak frequency $\nu_m$.  If we assume that the peak in
the synchrotron spectrum
passes through the optical band at the same time the light curve
breaks ($t\approx 700$ s),  as might be expected if the early optical light curve is dominated
by the internal shocks,
then we derive the parameters listed in Table 3.  As we describe below,
the pre-break optical light curve may also be due to external shocks,
in which case we
only have an upper limit on the passage time $t_{m,{\rm opt}}$ for the
synchrotron peak through the optical bands.  Aside from $\epsilon_B$,
the model parameters are sensitive
to (and roughly linearly proportional to) $t_{m,{\rm opt}}$ (Table 3).

\subsection{Optical/IR/UV Spectral Energy Distribution}
\label{sec:sed}

A tight constraint can be placed on the optical absorption considering only the observed optical/X-ray
spectral flux ratio $F_{\nu=R,{\rm obs}}/F_{\nu=\rm 1~keV}=467 \pm 2$, at $t=2\times 10^4$s in the observer frame.  
Here, the X-ray flux is corrected for absorption.  Allowing for the possibility that the synchrotron 
cooling break ($\nu_c$) is between observer frame $R$ band, 
$\nu_{R,{\rm obs}}$, and 1 keV in the observer frame, $\nu_{\rm 1~keV}$, the observed optical to X-ray flux is
\citep[e.g.,][]{galama01}
\begin{equation}
\label{equation:bbody}
F_{\nu=R,{\rm obs}}/F_{\nu=\rm 1~keV} = \left( { \nu_{\rm 1~keV} \over \nu_{R,{\rm obs}} } \right)^{1/2-\beta_{\rm opt}}
   \left( { \nu_c \over \nu_{R,{\rm obs}} }\right)^{-1/2} 10^{-0.4A_{R,{\rm obs}}},
\end{equation}
where $A_{R,{\rm obs}}$ is the absorption at $R$ band in the observer frame.  
The constraint on $\beta_{\rm opt}$ derived above for the constant-density model leads to $A_{R,{\rm obs}}<1.35$ mag.
For the wind density medium, there cannot be a spectral break due to the rapid X-ray versus optical light curve
decline, and we have $A_{R,{\rm obs}} = -6.8\beta_{\rm opt} - 6.7$.  This relation can only be satisfied with
positive absorption if $-\beta_{\rm opt} > 1$.  This is just possible in the wind scenario
for $\nu_c<\nu_{R,{\rm obs}}$, implying $p=1.6\pm 0.1$ \citep{chevNli00} and $A_{R,{\rm obs}} < 0.1$ mag.  
The rest-frame extinction in the $V$ band (i.e., 5556 \AA), $A_V$, will be a factor of a few times smaller than this, depending on
the extinction law.

\begin{figure}
\centerline{\includegraphics[width=7.0in]{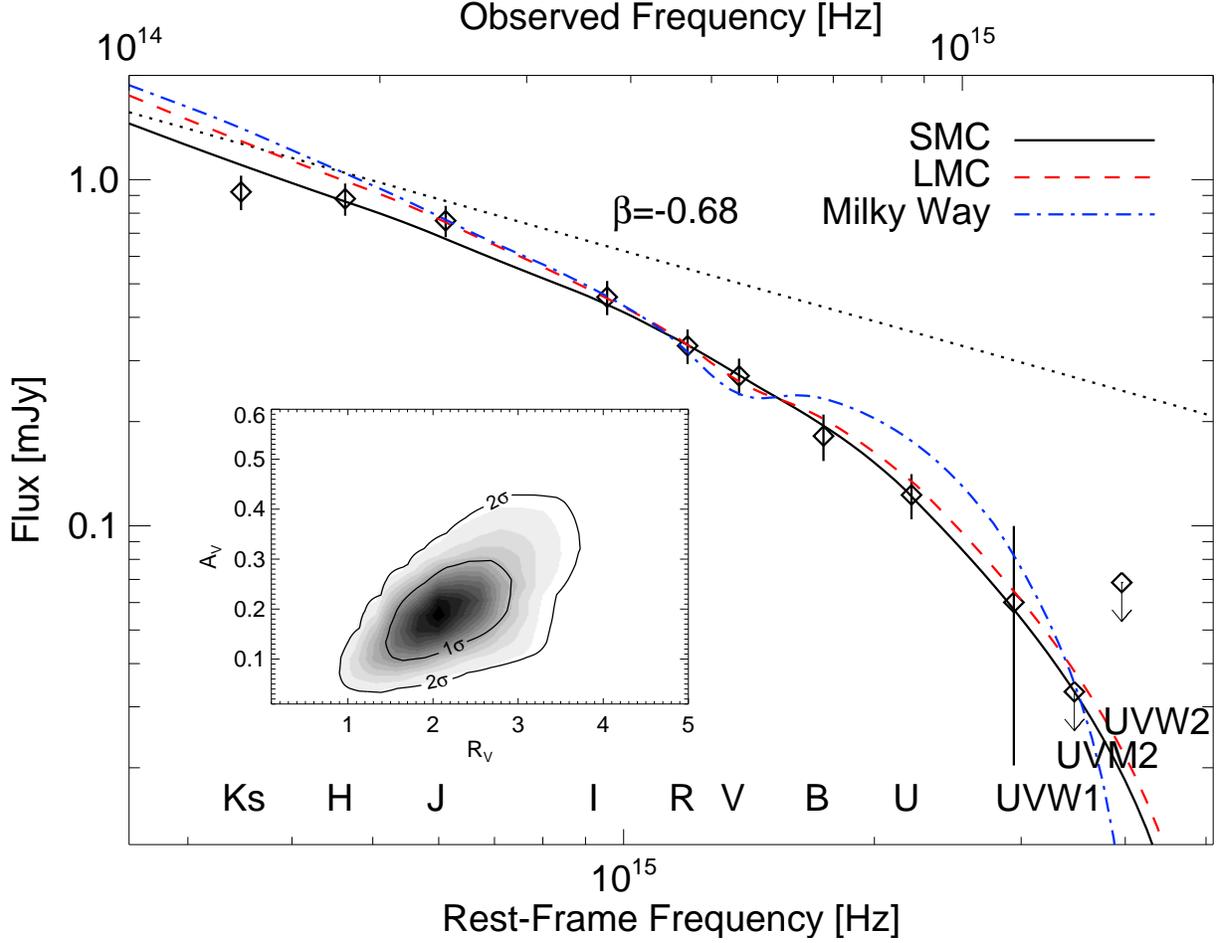}}
\caption{\small (a) Spectral energy distribution for the optical, IR, and UV data interpolated to $T=5200$s
after GRB~051111.  All data are corrected for Galactic
extinction \citep{sfd98}.  The dashed curve is a powerlaw with (energy) index $\beta = -0.68$, the best-fit value
found from the synchrotron shock modelling.  The other curves represent fits of the empirical extinction models from
\citet{pei92}, with the models labeled in the legend and discussed in Section \ref{sec:sed}.
(b; inset) The posterior probability for $A_V$ and $R_V$, using the \citet{reichart01} model prior.  The $1\sigma$ and $2\sigma$
confidence contours are overplotted.}
\label{fig:sed}
\end{figure}

Figure
\ref{fig:sed}a displays extinction-law fits to the optical/IR/UV spectral energy distribution (SED) interpolated to $5200$ s.
We interpolate all data using the broken powerlaw fit 
determined for the unfiltered KAIT and $R$ band Lulin data (Figure \ref{fig:lc_opt}).
We choose 5200 s so that little interpolation is required for the IR or UV data.
Because there is a prominent departure in the IR, optical, and UV light curve data from the fit
curve in the time interval $5\times 10^3 - 1.2\times 10^4$s in Figure \ref{fig:broadband_lc}
(see, also, Figure \ref{fig:lc_opt})---which can be modelled with a single magnitude offset of $\sim 0.6$ mag from the powerlaw curve
---we must include a variable offset in the SED fitting to describe the flux offset in this time interval.
(Explanations for this behavior are discussed below.)
Also, we use the constraint derived above for the (unabsorbed) optical energy index $\beta_{\rm opt}$.

We begin by fitting the empirical Milky Way, Large Magellanic Cloud (LMC), and Small Magellanic Cloud (SMC)
extinction curves of \citet{pei92}.  The fits (Figure \ref{fig:sed}a) yield
rest-frame $V$ band extinction values of
$A_V=0.38\pm 0.15$ ($\chi^2/\nu=12.23/7$), $0.35\pm 0.08$ ($\chi^2/\nu=5.96/7$), and
$0.23\pm 0.07$ ($\chi^2/\nu=4.32/7$) mag, respectively.  The excellent fit of the SMC model and the poor fit of the Galactic
extinction model point toward a weak 2175 \AA~dust ``bump'' and an increased level of far-UV extinction relative to that
found in the Galaxy.
Additionally, we fit the data using the general, 8-parameter family of
models from \citet{fm99}, focusing on variations in $A_V$ and the ratio of total to selective extinction,
$R_V = A_V / E(B-V)$.  In order to fold in prior knowledge of physically realistic values for
$R_V$ while not suppressing possible variations in the other parameters describing the 
extinction curve (e.g., the magnitude of the dust bump, which we marginalize over),
we exploit the \citet{reichart01} prior.  The 
best-fit extinction curve has $A_V=0.2\pm 0.1$ mag, in agreement with the SMC value determined above, and $R_V=2.0\pm 1.0$
($\chi^2/\nu=4.9/6$).  The model extinction at $R$ band in the observer frame is $A_{R,{\rm obs}}=0.7\pm0.1$ mag,
consistent with the value derived above from the optical/X-ray flux ratio alone.  Finally, we note that
the extinction curve of \citet{calz00} provides a poor fit ($\chi^2/\nu=9.42/6$, with best-fit $A_V=0.24$ mag, $R_V=2.0$).

Standard Galactic dust has $R_V= 3.1$ \citep{sneden78}, and
the low inferred $R_V$ \citep[see also, e.g.,][for the appearance of such
values in the SEDs of supernovae]{krisc00} implies a dust-grain 
distribution skewed toward fine grain sizes.  
The posterior probability contours in Figure \ref{fig:sed}b show $A_V<0.45$ mag (2$\sigma$).
Larger values of $A_V\approx \beta_{\rm opt} \approx 3$ mag are possible if we relax the constraint on $\beta_{\rm opt}$ and
allow positive values, however, these models are inconsistent with the temporal decay observed in the light curve
and cannot produce the observed optical/X-ray flux ratio.  
For the wind-density medium, which is excluded by the rapid X-ray versus slow optical flux decay unless the early
X-ray light curve is actually dominated by low-level flaring, the spectral slope can be as shallow as $\beta_{\rm opt}
=-0.3$ ($\nu_c$ above the X-ray band, in conflict with the constraint derived above from the optical/X-ray flux ratio).  
But this has little effect on the inferred $A_V$, and we find $A_V<0.5$ mag ($2\sigma$).

Both the SED fits and the optical/X-ray flux ratio demand a
considerably smaller dust column than we infer from the soft X-ray absorption (Section \ref{sec:xandg})
or from detailed modelling of the GRB~051111 optical spectrum.
The Galactic $N_H-A_V$ relation \citep{pns95} and the observed X-ray column imply $A_V=4\pm2$ mag.
(One caveat here is that XRT low energy calibration efforts are ongoing\footnote{http://swift.gsfc.nasa.gov/docs/swift/analysis/xrt\_digest.html}.)
This indicates that the ISM exhibits a low dust-to-gas ratio, similar to the SMC which is $\sim 8$
times smaller than Galactic \citep{bouchet85,pei92}.  Another contributing factor could be an overabundance
of the light metals \citep[which dominate the opacity at soft X-ray wavelengths;][]{mm83}, as also suggested by 
optical spectroscopy in the case of GRB~050401 \citep{watson06}.  
Similar to GRB~050401, the Keck/HIRES spectrum of GRB~051111 \citep{prochaska05,penprase06,prochaska06} exhibits strong absorption lines 
in the trace element Zn (implying $\log(N_{H,{\rm gas}}) = 21.2\pm 0.2$, for solar metalicity), which combines with
the inferred light metal column from the X-ray data to imply a factor of $4.0^{+4.0}_{-2.6}$ overabundance
in the multiple-$\alpha$ elements relative to the Fe-group elements.  Because of the saturation of the Zn absorption in
 the \citet{watson06} GRB~050401 spectrum
 and because Zn is only a trace element and does not contribute directly to the opacity (e.g., at X-ray wavelengths), this
possibility should be approached cautiously.  

In a separate paper, we plan to study in more detail the X-ray absorption in this 
burst relative to the absorption properties determined from the optical spectroscopy.
From the metal abundances inferred from the GRB~051111 Keck spectrum (e.g., Si/Fe or Zn/Fe),
we find evidence that the gas is highly depleted by dust.  
There is also marginal evidence that the light metal column (in gas form, $\log(N_{H,{\rm gas}}) = 21.0^{+0.7}_{-0.5}$,
from  $\log(N_{Si^+})= 16.6^{+0.7}_{-0.5}$) may be a factor $\sim 10$ smaller than that inferred from the X-ray
data.  It is not clear how these column densities are to be reconciled with the constraints imposed by the optical and
X-ray SED and the soft X-ray absorption.
The Keck spectrum shows, for example, $N_{Si^{++}} < N_{Si^+}$, which argues against an explanation in terms of a
significant ionization of the column.  The large X-ray column is probably not significantly due to intervening systems
(a MgII absorber in the optical spectrum at $z=1.189$ only contributes $\log(N_H)\approx 19.5$).

\subsection{External Shock Origin for the Early Optical Light Curve and GRB?}
\label{sec:broadband2}

From the optical colors, we derive above a stringent constraint on the
change in the spectral slope across the break
at $t\approx 700$ s, which excludes an explanation based on a passage through
the spectral bands of the synchrotron cooling break
\citep[yielding $\Delta \beta=0.5$; e.g.,][]{spn98}.  The most commonly invoked
cause of achromatic breaks---flux decrements due to finally viewing
the edge of the sideways expanding GRB jet \citep[e.g.,][]{rhoads99}---
is ruled out here by the mildness of the break ($\Delta \alpha = 0.19\pm
0.04$) and by the extremely slow optical fade after the break.
Another possible explanation, which grafts directly onto the $t>700$ s
solution discussed above, is that the density prior to the
break is increasing with radius.  A decreasing density, as for a wind medium
\citep[e.g.,][]{rykoff04},
does not work because the flux increase due to the decreasing optical absorption
is over-compensated by the decrease in synchrotron flux with radius.
We would also expect to see a strong color change due to the changing absorbing
column.  For the increasing density model, the absorbing matter close to the burst
is unimportant, and there is little expected evolution in the absorption with time.

The pre-break decay
is shallower by $\Delta \alpha = s/(8+2s)$, for a density rising as
$n\propto R^s$ and $s=4\pm 2$.  This is because the flux between the synchrotron peak frequency
and the cooling break is proportional to the synchrotron peak frequency $\nu_m^{(p-1)/2}$ times
the peak flux $F_m$ \citep[e.g.,][]{spn98}.  The frequency $\nu_m$ is independent of $s$, while
$F_m \propto t^{s/(8+2s)}$ \citep[e.g.,][]{chevNli00,spn98}.
The implied increase in density from $t\approx 30$ s to
$t\approx 10^3$ s is a factor of $\sim $5.  
Such a picture could potentially describe the prompt emission via the same external shock
that later generates the afterglow.  An external shock explanation for the prompt emission
is motivated superficially by
the apparent smoothness of the GRB \citep[see, e.g.,][]{mkp04}.  Minimally,
the model must be able to reproduce the extremely
hard GRB spectrum, with $-\beta_\gamma\approx 0.3-0.5$ extending to $E\approx
1$ MeV.  During the \citet{bm76} evolution stage when the fireball expands adiabatically, neither the synchrotron peak frequency 
($\nu_m$) nor the synchrotron cooling frequency ($\nu_c$) can decrease more rapidly than $t^{-3/2}$ and go
from such a high value at $t\approx 10$ s to the
derived values above at $t\approx 10^3$ s.  However, prior to this, during the fireball deceleration phase when the Lorentz
factor $\Gamma$ is expected to be roughly constant, the cooling frequency will decrease very rapidly,
$\nu_c\propto t^{-2-3s/2}$.  Our fits of this model, however, with an initial Lorentz factor
$\Gamma_{\circ}\approx 700$ \citep[e.g.,][]{mnr97} for $t<t_d = 10$ s (Figure \ref{fig:broadband_lc2}),
show that the external shocks which generate the optical flux underproduces gamma-rays by a factor $\sim 10$.
Synchrotron self-compton emission cannot help due to that mechanism's inefficiency at small $\epsilon_e$ \citep{sariesin01}.
A more careful calculation of the expected GRB flux from the \citet{bm76} solution \citep[e.g.,][]{granot99}, with
modifications to the standard synchrotron spectrum to possibly account for the harder GRB versus optical spectrum,
is beyond the scope of this paper.

\subsection{Energy Injection}

Refreshed shocks offer a plausible and perhaps better-travelled explanation \citep[e.g.,][]{nousek06} for the flat early light curve.
Figure  \ref{fig:broadband_lc2} shows the external shocks model discussed above, but with the addition of a changing shock
energy $E\propto t^a$ for $t<t_{\rm break}$.  For a slow-cooling, constant-density synchrotron model between $v_m$ and $v_c$,
the light curve drops as $t^{ a(p+3)/4 + \alpha }$ \citep{snm00}, where $\alpha=3(1-p)/4$ is the decay index without energy
injection \citep[e.g.,][]{spn98}.  We derive $a=0.20\pm 0.05$.  Between 60 and 700 s, the shock energy increases by $\sim 60$\%.
If the energy injection is due to a changing luminosity of the long-lasting central engine \citep{reesNmes00,macfad01,rr04,leeNrr02}, the
luminosity goes as $t^{-0.8\pm0.1}$, which may be challenging for the progenitor models \citep[e.g.,][]{macfad01}.  If instead the central
engine generated a flow with a powerlaw distribution of Lorentz factors $\Gamma$, with the slower shells of material
gradually catching up to the shock, we find $M(\Gamma) \propto \Gamma^{-1.6\pm 0.1}$ \citep{reesNmes98,snm00,rretal01}.

In the constant-density case, the optical flux is more sensitive to the
energy injection than is the X-ray flux \citep[by a factor $\Delta \alpha = a/4$; e.g., Equation (2) in][]{nousek06}.  If X-ray data
were available for GRB~051111 during the energy injection episode, we could therefore verify and possibly better test the scenario.
Long wavelength observations are particularly important for testing the claims of strong energy injection episodes ($\Delta \alpha \approx 1$),
inferred from the X-ray data alone \citep{nousek06}.

\subsection{The Optical Flux as Reprocessed Prompt Emission}

\begin{figure}
\centerline{\includegraphics[width=7.0in]{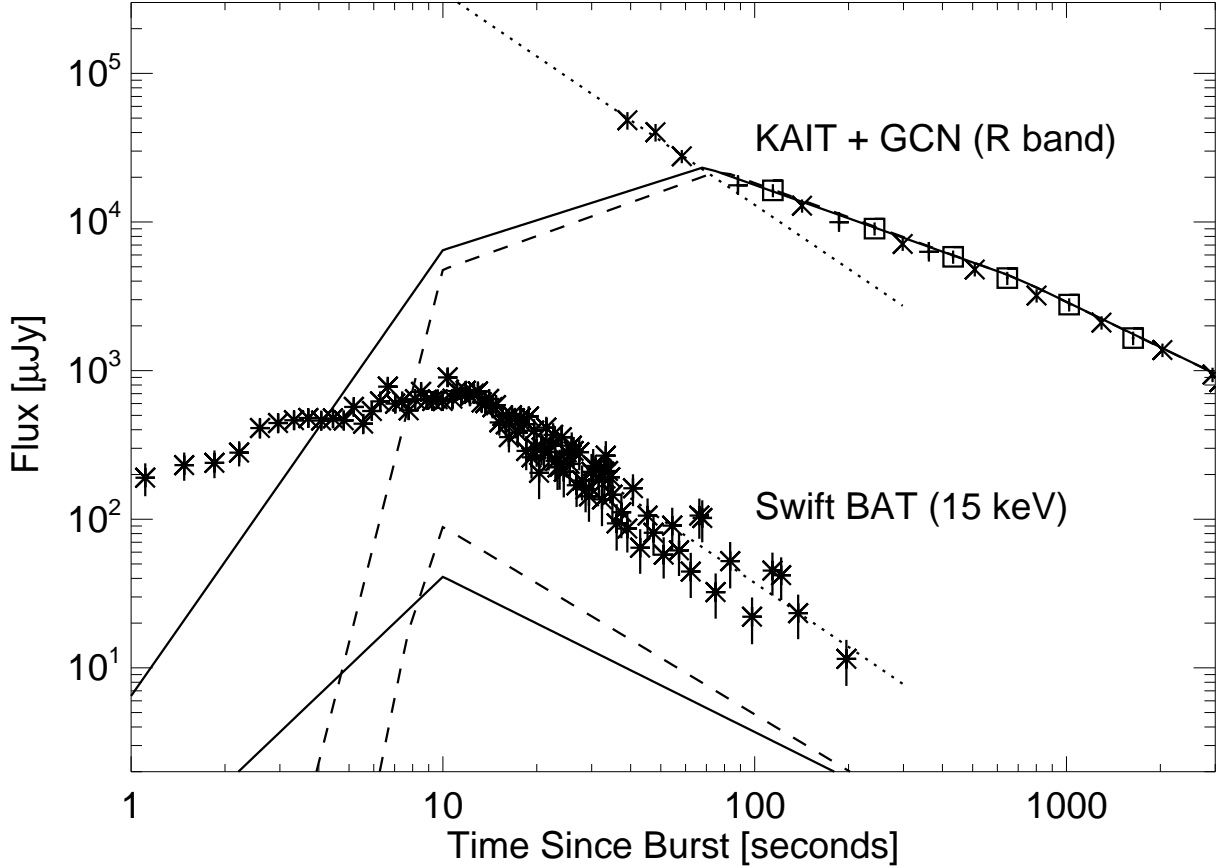}}
\caption{\small The early optical and prompt gamma-ray data fit with two variations on the
external shock model: energy injection prior to $t=700$ s with $E_{\rm shock} \propto t^{0.2}$ (solid lines),
and an
increasing density $n\propto R^4$ prior to $t=700$ s (dashed lines).  Due to an assumed constant initial Lorentz
factor $\Gamma_{\circ}=700$, all curves rise prior to the estimated deceleration time $t_d=10$ s.  Neither model
reproduces the strong gamma-ray flux.  The best-fit powerlaw for the prompt light curve decline ($F_{\gamma}\propto t^{-1.4\pm0.1}$)
is shown
as a dotted line, and also scaled by a factor of 350 to match the first three optical data points.
Here, we assume that the synchrotron peak frequency crosses the $R$ band
central frequency at $t_{\circ}=70$s (e.g., Table 3), which is an upper limit.  We include the GCN data
\citep{milne05, rujo05}, in addition to the data from KAIT and {\it Swift}. All data are corrected for Galactic
extinction \citep{sfd98}. The optical data are shown mapped in frequency to the $R$ band using the best-fit shock plus
absorption model (Table 3, $A_{R{\rm ~obs}}=0.72$ mag, Figure \ref{fig:sed}).}
\label{fig:broadband_lc2}
\end{figure}

Similar to what was found above for the increasing density model, the GRB flux at 15 keV in the energy injection scenarios
is $\sim 10$ times brighter than the expected
flux from the external shock (solid line in Figure  \ref{fig:broadband_lc2}).
There is a possible steepening
of the light curve decay near $t\approx 60$ s apparent from ROTSE \citep{rujo05}, Super-LOTIS \citep{milne05}, and the first KAIT
observation (Figure \ref{fig:lc_opt}).  This early flux decline is too steep, even assuming that energy injection has not yet begun
to occur at this epoch. These three data points therefore suggest that the earliest optical flux is dominated by                 
emission from the GRB.

The dotted curve in Figure  \ref{fig:broadband_lc2} shows a powerlaw fit to the gamma-ray data overplotted on the
optical.  The implied broadband slope is $\beta_{{\rm Opt}-\gamma} = -0.7$, consistent with the afterglow spectral slope
in the optical but marginally inconsistent with the prompt gamma-ray spectral slope (Section \ref{sec:xandg}).  If there is a time delay for
the reprocessing of the gamma-ray photons as has recently been proposed \citep{vestrand06}, this is a coincidence and the 
broadband slope could easily be consistent with the gamma-ray spectral slope.  In any case, for this and the rising density
model discussed above, the synchrotron peak frequency passes through the optical quite early ($t\lessim 60$ s), implying a very
inefficient transfer of the shock energy to the synchrotron emitting electrons (Table 3), which may also be true for the prompt emission.

\subsection{Light Curve Variability}

The optical light curve at $t\gtrsim 10^4$ s (subpanel of Figure \ref{fig:lc_opt2}; Section \ref{sec:nocolor}) and the optical/IR/UV light
curve in the range $5\times 10^3 - 1.2\times 10^4$ s show evidence for residual variability at the $\gtrsim 30$\% level,
with $dt/t\approx 0.1$.  There is little evidence for spectral change during the variability.
Such variability may be common and has been seen previously for well sampled GRB afterglows.
The light curve of GRB~021004 \citep{shira02} displayed several prominent bumps \citep{bersier03,mirabal03,fox03a}.
The exquisitely sampled light curve of GRB~030329 \citep{lipkin04,van03} displayed prominent departures from a broken powerlaw
fade, even when the underlying supernova emission was subtracted away \citep{bersier03}.

\citet{npg03} explore several possible explanations for the variability in GRB~021004 and find that an explanation in
terms of in terms of refreshed shocks, a non-uniform (or ``patchy'') GRB jet, or variations in the external density
\citep[see, also,][]{wangNloeb00,daiNlu00,laz021004} are all possible.  \citet{gnp03} carry out a similar analysis
for GRB~030329 and suggest that density variations are unlikely to be responsible for the optical variability due to the
apparent passage of the synchrotron cooling break through the passband \citep[e.g.,][]{bersier03}.  Late-time variability
after an apparent jet break also argues against the patchy jet model, although that model could still describe the
variability observed at earlier times.

For GRB~051111, the variability occurs during the phase ($t\gtrsim 700$ s) when the afterglow appears to be well described by external shocks.  
We believe that this favors an explanation in terms of
density variation in the surrounding medium.  Because the X-ray data are not
expected to be affected by small variations in the density, this explanation is backed up
by the quality of the X-ray fit with a single temporal powerlaw, although the X-ray error bars are $\sim 30$\%.
Recently, \citet{guido05} argue that density enhancements can explain an
achromatic optical light curve bump in data taken 3 minutes after GRB~050502A.
More finely sampled data, with broad spectral coverage, will be required to pin down the true source of this variability in GRB afterglows.

\section{Conclusions}

We have presented a thorough analysis of the rich broadband data available 
for {\it Swift} GRB~051111, with a focus on the early, multi-band optical data
from KAIT and Lulin.  The optical data prior to $t\approx 700$ s show a very flat
decline, with little or no evidence for a color change as compared to the data
after $t\approx 700$ s.   The data at $t\gtrsim 1$
minute to several hours after the GRB are well fit using a simple, modified external shock
model with absorption by gas and dust.  The modelling entails energy injection or a rapidly increasing density profile
with radius, prior to the break time.  At later times there
are large ($\gtrsim 30$\%), possibly achromatic modulations in the optical, IR, and UV about the best-fit model.
The modulations appear to be common in well-studied optical afterglows, yet their origin remains mysterious.

The increasing density model may
allow for an external shock explanation of both the prompt gamma-ray and later emission.  Such an increase in density might
be expected if a supernova (SN) occurred prior to the GRB, as in the ``supra-nova'' model \citep{vietri99}.  However,
the simultaneity of GRB and SN in the nearby, well-studied cases (GRB~060218/SN~2006aj, GRB~980425/SN~1998bw, GRB~031203/SN~2003lw)
argues against this possibility.   

On the other hand, slowly declining light curves in the X-ray band are common and are thought to
be due to shock refreshment \citep[e.g.,][]{nousek06}, and the GRB~051111 observations from KAIT show that this emission can
also dominate the optical light curve at early times.  In fact, the optical light curve for this event, which shows a rapid early
decline, followed by a levelling off and then a moderate decline typical of those found in the past and modelled with external
shocks, appears quite similar to the ``canonical'' behavior observed in the X-ray band and reported by \citet{nousek06}.
Perhaps there is a canonical optical afterglow behavior, too.  (As a counter-point, the early optical behavior reported for GRBs
060206 \citep{monfar06,stanek06} and 060210 \citep{stanek06}
appears quite different from that here.)  

An important feature of the GRB~051111 optical afterglow is the
lack of a turnover in the optical decay rate at early times.  We do not detect the peak in
the synchrotron spectrum passing through the optical bands.  Consonant with
a lower than average X-ray flux at late times, this leads to
a low value for the fraction (assuming constant equipartition) of
shock kinetic energy winding up in the synchrotron-emitting electrons, $\epsilon_e \lessim 0.3$\%.
Such a low value for $\epsilon_e$ is uncommon, but not unheard of, in GRBs \citep[see, e.g.,][]{pnk02}.
The very low value may indicate a wider diversity than previously suspected in the microphysical parameters from GRB to GRB \citep[see, also,][]{bkf03}.
Aside from the low $\epsilon_e$ value, the derived external shocks parameters are comparable to
those previously found.

It is important to stress that whether we find $\epsilon_e \approx 0.3$\% or $\approx 3$\% depends on the deconvolution
of the early data into prompt internal shock and afterglow external shock components.
Our favored smaller $\epsilon_e$ arises if the synhrotron peak frequency has passed through the optical well before
the break at $t\approx 700$ s, in which case mild energy re-injection into the shock can explain the gradually decaying light curve.
To get the larger $\epsilon_e$ value, the light curve prior to $t\sim 700$ s must be dominated by the GRB and not by the
external shocks which dominate after $t\sim 700$ s.  Otherwise, the early light curve would be rising rather than
declining.  In any case, reverberations of the prompt emission reprocessed into optical
radiation \citep[e.g.,][]{vestrand06} appear to be required to explain the earliest few optical data points for GRB~051111.
It does not appear to be necessary to invoke reverse-shock emission \citep[e.g.,][]{akerlof99,li2003b}, which is expected
to produce time decays more rapid than those observed.

The absorption at IR, optical, and UV wavelengths in the observer frame
is well fit by an SMC extinction curve.   Little evidence for the 2175 \AA~dust ``bump'' and an excellent fit 
of the SMC extinction profile are common features in
optical GRB afterglow spectra \citep[e.g.,][]{vrees04,jakob03,sNf04,watson06}.  Combined with the soft X-ray
absorption measurement, there is an implied low dust-to-gas ratio and
a possible overabundance of the light-metals relative to the Fe-group metals.
This has also been observed for GRB~050401 \citep{watson06}.
The light metal overabundance works against a direct association with an SMC-like
environment, because the SMC has a metalicity $\sim 1/10$ times solar \citep{pei92}.

The SED fitting also implies a low ratio of total to selective extinction, $R_V\approx 2$.
This is a clue that the absorbing medium exhibits unusual dust properties.  The
work of \citet{galama01} establishing large typical $N_H/A_V$ values has
led to suspicions that
GRBs could destroy dust out to distances $R\approx 20$ pc \citep[see also,][]{wNd00,fruchter01,dNh02,pNl02,perna03}. 
The small dust grains are preferentially destroyed, creating a flat (or ``gray'') extinction 
curve \citep{galama01,sNf04,stratta05}.  A small $R_V$, however, suggests small dust grains are dominant, and our SED is very curved.
These facts argue against the GRB playing a direct role in defining the extinction properties.
The unchanging optical
color implies that the absorbing column is not local to the GRB (i.e., $R\gtrsim 0.1$ pc) and may be
associated with a nearby giant molecular cloud or the GRB host galaxy.
Time-resolved spectroscopy in the optical and X-ray bands is of utmost importance
for answering these questions.

For GRB~051111, as for a few other {\it Swift}~events for which ground-based observers
have been fortunate enough to capture early data, there is an emerging complicated interplay between
the GRB and the subsequent shocking of the external medium.  It is critical that more long-wavelength
data be taken and published for other GRBs in order to facilitate modelling similar to that performed above.
 Only with such a disentangling of the competing emission processes
can we hope to answer open questions regarding which shock components
are truly the most important and what microphysical parameters define the shocks and characterize the
transition from internal to external shocks.

\acknowledgments
N. Butler gratefully acknowledges support from a Townes Fellowship at the U.~C. Berkeley 
Space Sciences Laboratory, as well as partial support
from J.S.B and A.V.F.   The work of A.V.F.'s group
is supported by NASA/{\it Swift} grants NNG05GF35G and NNG06GI86G.
J.S.B., J.X.P., and H.-W.C. are partially supported by NASA/{\it Swift}
grant NNG05GF55G.
KAIT and its ongoing research were made possible by generous donations from Sun
Microsystems, Inc., the Hewlett-Packard Company, AutoScope Corporation, Lick
Observatory, the National Science Foundation, the University of California,
the Sylvia \& Jim Katzman Foundation, and the TABASGO Foundation.
This work is partly supported by grants NSC 94-2752-M-008-001-PAE, NSC
94-2112-M-008-002, and NSC 94-2112-M-008-019.
Y.U acknowledges support from the Japan Society for the Promotion of
Science (JSPS) through JSPS Research Fellowships for Young Scientists.

\vfill
\eject

\input tab1.tex
\input tab2.tex
\input tab3.tex

\end{document}

%% file: tab1.tex
\begin{table}
\begin{center}
\caption{KAIT Photometry of GRB~051111}
\vspace{5mm}
\footnotesize
\begin{tabular}{ccccc}\hline\hline
 $t$ [s] & Exp. [s] & Mag. & $\sigma_{\rm Mag.}$ & Band \\\hline
43.7 & 15 & 13.60 & 0.02 & R \\
73.7 & 15 & 14.54 & 0.02 & V \\
99.7 & 15 & 13.57 & 0.02 & I \\
124.7 & 20 & 14.43 & 0.02 & R \\
156.7 & 45 & 15.16 & 0.03 & V \\
212.7 & 45 & 14.21 & 0.02 & I \\
268.7 & 45 & 15.08 & 0.02 & R \\
324.7 & 60 & 15.66 & 0.02 & V \\
395.7 & 60 & 14.69 & 0.02 & I \\
470.7 & 60 & 15.51 & 0.02 & R \\
578.7 & 120 & 15.04 & 0.02 & I \\
734.7 & 120 & 15.94 & 0.02 & R \\
892.7 & 240 & 15.49 & 0.02 & I \\
1167.7 & 240 & 16.40 & 0.03 & R \\
1448.7 & 360 & 16.05 & 0.03 & I \\
1843.7 & 360 & 16.86 & 0.03 & R \\
2873.7 & 120 & 17.27 & 0.05 & R \\
3026.7 & 120 & 17.40 & 0.06 & R \\
3178.7 & 120 & 17.36 & 0.05 & R \\
3330.7 & 120 & 17.41 & 0.06 & R \\
3486.7 & 120 & 17.49 & 0.06 & R \\
3639.7 & 120 & 17.46 & 0.07 & R \\
3791.7 & 120 & 17.51 & 0.06 & R \\
3944.7 & 120 & 17.62 & 0.07 & R \\\hline
\end{tabular}
\end{center}
{\small Notes: $R$ band magnitudes determined for unfiltered observations (Section \ref{sec:redux}).
We denote with $t$ the elapsed time since the BAT trigger.}
\label{tab:051111_kait}
\end{table}

%% file: tab2.tex
\begin{table}
\begin{center}
\caption{Lulin Photometry of GRB~051111}
\vspace{5mm}
\footnotesize
\begin{tabular}{ccccc}\hline\hline
 $t$ [hr] & Exp. [s] & Mag. & $\sigma_{\rm Mag.}$ & Band\\\hline
4.2201 & 300 & 19.52 & 0.05 & V \\
4.3154 & 300 & 19.12 & 0.04 & R \\
4.4107 & 300 & 18.45 & 0.04 & I \\
4.5071 & 300 & 20.38 & 0.07 & B \\
4.6021 & 300 & 19.56 & 0.05 & V \\
4.6974 & 300 & 19.15 & 0.03 & R \\
4.8012 & 300 & 18.51 & 0.04 & I \\
4.9154 & 300 & 20.30 & 0.07 & B \\
5.0085 & 300 & 20.37 & 0.08 & B \\
5.1018 & 300 & 20.25 & 0.07 & B \\
5.3204 & 300 & 20.32 & 0.07 & B \\
5.4157 & 300 & 19.54 & 0.04 & V \\
5.5110 & 300 & 19.20 & 0.03 & R \\
6.0479 & 300 & 19.76 & 0.05 & V \\
6.1435 & 300 & 19.27 & 0.04 & R \\
6.2387 & 300 & 18.59 & 0.04 & I \\
6.4037 & 300 & 20.55 & 0.06 & B \\
6.4990 & 300 & 19.89 & 0.06 & V \\
6.6893 & 300 & 19.73 & 0.05 & V \\
6.7846 & 300 & 20.88 & 0.08 & B \\
6.8796 & 300 & 19.86 & 0.05 & V \\
6.9813 & 300 & 19.47 & 0.05 & R \\
7.0762 & 300 & 18.68 & 0.03 & I \\
7.1799 & 300 & 20.86 & 0.08 & B \\
7.2729 & 300 & 20.73 & 0.10 & B \\
7.3662 & 300 & 20.88 & 0.10 & B \\
7.5546 & 300 & 19.94 & 0.07 & V \\
7.6479 & 300 & 19.96 & 0.05 & V \\
7.7926 & 300 & 19.50 & 0.03 & R \\
7.8879 & 300 & 18.78 & 0.04 & I \\
8.0488 & 300 & 20.72 & 0.08 & B \\
8.1421 & 300 & 21.01 & 0.14 & B \\
8.2365 & 300 & 21.03 & 0.14 & B \\
8.3318 & 300 & 20.25 & 0.10 & V \\
8.4257 & 300 & 20.23 & 0.08 & V \\
8.5193 & 300 & 20.18 & 0.08 & V \\
8.6235 & 300 & 19.82 & 0.06 & R \\
8.7190 & 300 & 19.21 & 0.06 & I \\\hline
\end{tabular}
\end{center}
{\small Notes: We denote with $t$ the elapsed time since the BAT trigger.}
\label{tab:051111_lot}
\end{table}

%% file: tab3.tex
\begin{table}
\begin{center}
\caption{Broadband Fit Parameters, Constant-Density ISM}
\vspace{5mm}
\small
\begin{tabular}{cc}\hline\hline
 Paramater & Value \\\hline
$E_{\gamma, \rm iso}$ &  $6.2 \times 10^{52}$ erg (fixed) \\
$D_{\rm lum}$ & $3.46 \times 10^{28}$ cm (fixed) \\
$z$ & 1.55 (fixed) \\
$\eta_{\gamma}$ &  $(2.1\pm0.3) ~(t_{m,{\rm opt}}/[700~s])^{1.39\pm 0.05}$ \\
$p$ & $2.35\pm0.05$ \\
$\epsilon_e$ & $(0.03\pm0.01) ~(t_{m,{\rm opt}}/[700~s])^{1.12\pm 0.05}$ \\
$\epsilon_B$ & $(0.02 \pm 0.01) ~(t_{m,{\rm opt}}/[700~s])^{-0.11\pm 0.15}$ \\
$n$  & $(0.8\pm 0.5) ~(t_{m,{\rm opt}}/[700~s])^{0.86\pm 0.12}$ cm$^{-3}$ \\\hline
\end{tabular}
\end{center}
{\small Notes: The fit in Figure \ref{fig:broadband_lc} uses $t_{m,{\rm opt}}=700$ s for the passage of the synchrotron peak
frequency through the optical.}
\label{tab:broadband_fit}
\end{table}